\documentclass[prl,twocolumn]{revtex4}
\usepackage{graphicx}% Include figure files

\begin{document}

%\draft

\title
{\large{\bf The frustrated Brownian motion of nonlocal solitary waves}}
\author{V. Folli$^1$, C. Conti$^{2}$}
\affiliation{
$^1$Department of Physics, University Sapienza, Piazzale Aldo Moro 2, 00185, Rome (IT)\\
$^2$Institute for Complex Systems (ISC-CNR), Department of Physics, University Sapienza, Piazzale Aldo Moro 2, 00185, Rome (IT)}
\email{claudio.conti@roma1.infn.it}
\date{\today}

\begin{abstract}
We investigate the evolution of solitary waves in 
a nonlocal medium in the presence of disorder.
By using a perturbational approach, we show that an increasing degree 
of nonlocality may largely hamper the Brownian motion of
self-trapped wave-packets.
The result is valid for any kind of nonlocality and in the presence of non-paraxial effects.
Analytical predictions are compared with numerical simulations
based on stochastic partial differential equations.
\end{abstract}

%\pacs{}

\maketitle
Wave-packets may display particle-like behavior in the presence of 
non-linearity. Solitary waves (SW) and solitons are the non spreading
solutions of the relevant nonlinear wave equations that describe
such a phenomenon. These self-trapped beams have been observed in a variety of physical
systems, ranging from oceanic waves to Bose Einstein condensates (BEC) \cite{KivsharBook,WhithamBook}.
Over the years the role of a nonlocal nonlinear response, with special emphasis
on the optical spatial solitons (OSS) \cite{TrilloBook}, appeared with
an increasing degree of importance \cite{Snyder97, Conti03,Rotschild05, Rasmussen05, Buccoliero07, Kartashov07, Ouyang09};
on one hand because it must be taken into account for the quantitative description of experiments and, on the other hand,
because it is a leading mechanism for stabilizing multidimensional solitons \cite{kroli00}.
Nonlocality in nonlinear wave propagation  is found 
in those physical systems exhibiting long range correlations,
like nematic liquid crystals (LC) \cite{Conti03}, photorefractive media (PR) \cite{Segev92},
thermal \cite{Rotschild05,Ghofraniha07,Conti09} and thermo-diffusive \cite{Ghofraniha09} nonlinear susceptibilities, soft-colloidal matter (SM) \cite{Conti05PRL}, BEC \cite{Parola98,Perez00}, and plasma-physics \cite{Litvak78,Pecseli80}.
When nonlocality is known to play a role, 
randomness is typically present and results from the unavoidable material fluctuations, and its role is fundamental for all-optical logic gates and soliton driven devices \cite{Peccianti04nature,Peccianti02}.

In recent years, widespread investigations dealt with the interplay
between randomness and nonlinearity, with emphasis on Anderson localization and SW
\cite{Skipetrov03, Conti07, Swartz07,Staliunas03,Sacha09,Fort05,Kartashov08}.
Understanding the interplay between nonlocality and randomness is 
hence a fundamental subject in the theory of self-trapped waves.
However, while the literature on solitons in random media is vast (see, e.g., \cite{Abdullaev05} and references therein), 
stochastic nonlocal models were considered only in a few works \cite{Conti2005PRE,Conti06}  and a quantitative theory for the SW fluctuations was not reported.

In this Letter, we consider the effect of randomness on the stochastic dynamics
of a nonlocal SW and show that, as the degree of nonlocality increases,
the amount of fluctuations diminishes, 
and vanishes for an infinite-range nonlocal response. 
This is somehow a counter-intuitive result because in such media 
the nonlinear perturbation extends far beyond the spatial extension of 
the self-trapped wave, which is hence expected to be influenced by the material fluctuations on
a scale much larger than in the local case.

The model we consider is written as 
\begin{eqnarray}
\label{psieq}
i \partial_t \psi+\nabla_x^2 \psi +\rho \psi=0\\
\label{rhoeq}
\mathcal{G}(\rho)=|\psi|^2+\eta(x,t)
\end{eqnarray}
where $\psi$ is the relevant wave field, $x$ is the position vector, $\mathcal{G}$ is a linear differential operator, which does not include derivatives with respect to the evolution coordinate $t$, and $\eta(x,t)$ is a Langevin noise 
 such that $\langle \eta(x,t) \eta(x',t')\rangle=\eta_N^2 \delta(x-x') \delta(t-t')$. The origin of $\eta$ 
depends on the specific physical problem: 
(i) temperature (nematic director) $\rho$ fluctuations for thermal (LC) media; 
(ii) SM particle density $\rho$ fluctuations;
(iii) space-charge field $\rho$ fluctuations for PR (eventually induced by modulation of the background field); (iv) finite-temperature results in terms like $\eta$ for plasmas and BEC \cite{Stoof99}.

In the Fourier domain (\ref{rhoeq}) is written as 
$\tilde\rho=S(q)(\tilde{|\psi|^2}+\tilde \eta)$, where the tilde denotes the Fourier
transform, $S(q)$ is the ``structure factor,'' \cite{Conti05PRL}, and the corresponding Green function is denoted by $K(x)$, such
that 
\begin{equation}
\label{psieq2}
i \partial_t \psi+\nabla_x^2 \psi +V(x,t) \psi+\psi K*|\psi|^2 =0\text{,}
\end{equation}
where $V(x,t)=K*\eta$ is a colored random noise (the asterisk ``*'' denoting the $x-$convolution integral).
The local regime corresponds to $K(x)= \delta(x)$, while in the highly nonlocal regime $K(x)=K_0$ \cite{Snyder97}.
Let $\phi=\psi\exp(i\beta t)$, Eq.(\ref{psieq2}) is written as 
\begin{equation}
\label{psieqpertu}
i \partial_t \phi+\nabla_x^2 \phi-\beta \phi+\phi K*|\phi|^2 =i s(x,\psi,\psi_x,\psi_{xx},...,t)
\end{equation}
where $s$ is taken as a perturbation term, depending on $\psi$ and its transverse derivative at any order, and $\beta$ is the nonlinear wave-vector. 
Eq.(\ref{psieqpertu}) is a generalization of (\ref{psieq2}), accounting for any kind of perturbation, e.g., in the presence of material
losses $s=-\alpha\phi$, with $\alpha$ the loss coefficient. Eq. (\ref{psieq2}) corresponds to $s=V \phi$.

Soliton perturbation theory was previously developed 
for one dimensional (1D) solitons of the integrable local nonlinear Schroedinger (NLS) equation (see,e.g., \cite{IannoneBook}), and it is based 
on the knowledge of the exact soliton solutions. Here this approach 
is generalized to a non-integrable model, in the presence 
of an arbitrary nonlocality, and then applied to the SW Brownian motion. 
These are given, in the absence of the perturbation ($s=0$), 
by the real valued solutions $u(x)$ of 
\begin{equation}
\label{psieq2bound}
-\beta u+\nabla_x^2 u +u \int K(x-x') u^2 (x') dx'=0\text{.}
\end{equation}
In absence of perturbation, the general solution of (\ref{psieqpertu}) is dependent on $\beta$ and
on a given number of parameters, due to the Lie symmetries of the model. 
To simplify the notation, we consider hereafter the 1D case,
while the results below equally apply to the multi-dimensional case. 
The un-perturbed solution is written as
\begin{equation}
\phi_0=u(x-X+2\Omega t,\beta)\exp(i\theta-i\Omega x-i \Omega^2 t)
\end{equation}
where translational, gauge and Galilean invariance are taken into account:
$X$ is the center of the self-trapped wave, $\theta$ is the phase, and $\Omega$ is the momentum.
Due to the Galilean invariance, the analysis can be limited to SW with zero velocity ($\Omega=0$).
By letting $\phi=\phi_0+\phi_1 $, the linearized evolution equation is 
\begin{equation}
 \partial_t \phi_1=\mathcal{L}(\phi_1)+s\text{,}
\label{linearized}
\end{equation}
with
\begin{equation}
\begin{array}{l}
\mathcal{L}(\phi_1)=-i\beta \phi_1 +i\phi_{1,xx} +\\
i\phi_0 K*(\phi_0 \phi_1^*+\phi_0^* \phi_1)+i\phi_1 K*|\phi_0|^2\text{.}
\end{array}
\end{equation}
By introducing the scalar product $(a,b)=\Re \int a^* b \,dx$, it can be readily 
verified that $\mathcal{L}$ satisfies the following relation 
$(a,\mathcal{L}(b))=(\hat \mathcal{L} (a),b)$ 
with $\hat \mathcal{L}$ the adjoint operator 
\begin{equation}
\begin{array}{l}
\hat\mathcal{L}(\phi_1)=i\beta \phi_1 -i\phi_{1,xx} -i\phi_0 \times \\
K*(\phi_0 \phi_1^*-\phi_0^* \phi_1)-i\phi_1 K*|\phi_0|^2
\end{array}
\end{equation}
and such that $\hat\mathcal{L}(i a)=-i\mathcal{L}(a)$. 
Without loss of generality, the first order perturbation 
can be decomposed in a term representing a small variation of the solitary-wave parameters and the remaining part, denoted as the radiation term $\phi_r$. The former is proportional to 
the derivatives of $\phi_0$ with respect to the various parameters, 
$X$,$\Omega$,$\theta$,$\beta$ , and the expression for $\phi_1$ is written as
\begin{equation}
%\begin{array}{l}
\phi_1=f_X \delta X +f_\theta \delta \theta +f_\beta \delta\beta+(f_{\Omega}-X f_\theta)\delta\Omega+\phi_r\text{,}
%\end{array}
\end{equation}
while having introduced the auxiliary functions
\begin{equation}
\begin{array}{l}
f_\theta=i\phi_0\\
f_\beta=\partial_\beta\phi_0\\
f_{X}=\partial_{X}\phi_0\\
f_{\Omega}=-i(x-X)\phi_0\text{,}
\end{array}
\end{equation}
and $\delta X(t)$, $\delta \theta(t)$, $\delta \beta(t)$ and $\delta \Omega(t)$ being the time-dependent perturbations
to the bound state parameters.
Any of the auxiliary function $f$ is such that $\partial_t f=\mathcal{L}(f)$, the adjoint functions $\hat{f}$ are defined by $\partial_t (\hat{f},f)=0$. They are given by $\hat{f}_\theta=i f_\beta$,$\hat {f}_\beta=-i f_\theta$,
$\hat{f}_{\Omega}=-i f_X$,
$\hat{f}_{X}=i f_{\Omega}$  and are such that
$(\hat{f}_a,f_b)=\mathcal{N}_a \delta_{a,b}$ 
with $a$ and $b$ two symbols in the ensemble 
($X$,$\Omega$,$\theta$,$\beta$).
By direct integration it turns out that 
$\mathcal{N}_\theta=\mathcal{N}_\beta=(1/2)(dP/d\beta)$ and
$\mathcal{N}_{X}=\mathcal{N}_{\Omega}=(1/2)P$, with $P$ is the 
propagation invariant SW norm, or power, $P=\int |\phi_0|^2 d\mathbf{x}=P(\beta)$. $dP/d\beta\equiv P'>0$ due to nonlocal SW stability \cite{KivsharBook}.
\noindent Following the original argument in \cite{Gordon86},
the functions $f$ and $\hat{f}$ are localized around the SW position $X$,
as a consequence, their scalar products with $\phi_r$ vanish
because the radiation term spreads for long times, and $(\hat{f},\phi_r)=0$ is invariant.
This assumption holds true at the lowest order of approximation in $s$,
as based on the integrable NLS equation, 
and is confirmed in the non-integrable case considered here by the agreement with numerical simulations reported below.

By using the previous expressions in (\ref{linearized}) and projecting on the
adjoint functions,  the following equations 
are derived for the dynamics of the SW parameters 
\begin{equation}
\label{ODE}
\begin{array}{l}
\delta\dot\theta-X \delta\dot\Omega =\delta\beta+
\displaystyle\frac{2 S_\theta}{P'}\text{,}\qquad
\delta\dot{\beta}=\displaystyle\frac{2 S_\beta}{P'}\text{,}\\
\delta\dot{X}=-2\delta\Omega+\displaystyle\frac{2 S_X}{P}\text{,}\qquad\delta\dot{\Omega}=\displaystyle\frac{2 S_\Omega}{P}\text{,}\\
\end{array}
\end{equation}
where $S_\alpha=\left( \hat{f}_\alpha,s\right)$, and the dot is the $t-$derivative.
\noindent Eqs. (\ref{ODE}) hold for any $s$; for a random perturbation in the 
density $\rho$, as given by $\eta(x,t)$ above, they become 
\begin{equation}
\begin{array}{l}
\delta\dot\theta=X\delta\Omega+\delta\beta+\displaystyle\frac{1}{P'}\frac{d}{d\beta}\int u^2(x-X)(K*f) dx\text{,}\\
\delta\dot\beta=0\text{,}\qquad
\delta\dot X=-2\Omega\text{,}\\
\delta\dot \Omega=-\displaystyle\frac{2}{P}\int u(x-X)u_x(x-X)(K*f) dx\text{,}
\end{array}
\end{equation} 
from which
\begin{equation}
\begin{array}{l}
\delta\Omega(t)=-\displaystyle\frac{2}{P} \int_0^t \\
\int\int u(x-X) u_x(x-X) K(x-x') f(x',t') dx' dx dt'\text{.}
\end{array}
\label{deltaOeq}
\end{equation}
By writing $\delta\Omega(t)\delta\Omega(t')$ after (\ref{deltaOeq}) and  averaging over disorder leads to
\begin{equation}
\langle \delta\Omega(t)\delta\Omega(t')\rangle=\displaystyle\frac{4\langle f\rangle}{P^2}C \min(t,t')
\end{equation}
where
\begin{equation}
\begin{array}{l}
C=\int\int\int u(x_1-X) u(x_2-X) u(x_3-X)\times\\
 u_x(x_3-X) K(x_1-x_2) K(x_3-x_2) dx_1 dx_2 dx_3
\end{array}
\end{equation}

\begin{figure}
\includegraphics[width=0.50\textwidth]{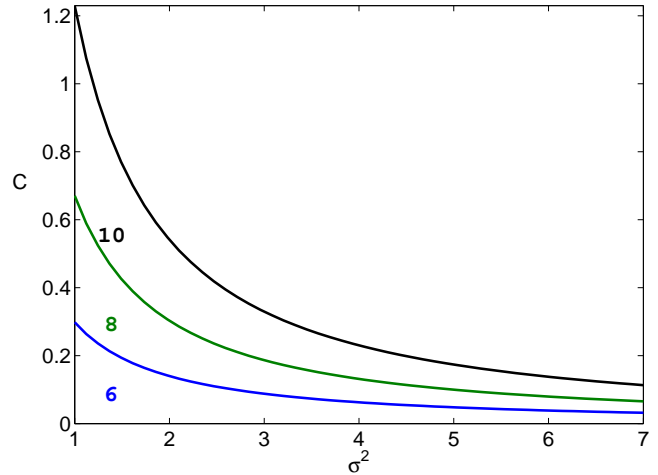}
\caption{(Color online)
Amplitude of the fluctuations $C$ for the exponential nonlocality versus $\sigma^2$, after Eq.(\ref{ghnonlocal}), for various soliton powers $P$.}
\label{fig1} \end{figure}

The deviation from the mean position is found as
\begin{equation}
\delta X(t)=-2 \int_0^t \delta\Omega(t') dt'
\end{equation}
from which $\langle\delta X\rangle=0$ and 
\begin{equation}
\langle \delta X(t)^2 \rangle=4 \int_0^t \int_0^t \langle 
\Omega(t_1)\Omega(t_2)\rangle dt_1 dt_2=\displaystyle\frac{16\eta_N^2 C}{3 P^2}t^3\text{.}
\label{ghnonlocal}
\end{equation} 
The previous result is the nonlocal counterpart of the so-called Gordon-Haus effect, first introduced for describing the 
random fluctuations of solitons in amplified light-wave systems \cite{Gordon86,IannoneBook}.
Eq.(\ref{ghnonlocal}) states that the random fluctuations,
measured by $\langle \delta X(t)^2 \rangle$, grow with the cubic of the propagation distance, and decay with the square of the 
SW power. A significant role is played by quantity $C$, which, in general, depends on the specific soliton and nonlocality profile. In the local regime $K(x-y)=\delta(x-y)$ it is 
\begin{equation}
C=\int \left[ u(x_1-X) u_x(x_1-X) \right]^2 dx_1\text{.}
\end{equation}
The relevant result is found in the highly nonlocal limit $K(x)\rightarrow K_0$, which gives,
for a bell-shaped soliton profile [$u(x)=u(-x)$],
\begin{equation}
C=K_0^2 \left[\int u(x_1-X) u_x(x_1-X) dx_1\right]^2=0\text{,}
\end{equation}
irrespective of the specific shape of $K(x)$.
As a result,  {\it in the highly nonlocal regime the random fluctuations of the fundamental soliton vanish.}
Physically, this corresponds to the fact that spectral power density of the noise is averaged out by a narrow $S(q)$ as the degree of nonlocality increases.
We stress that this results is independent of the specific kind of nonlocality.
For example, with reference to the exponential nonlocality $S(q)=(1+\sigma^2 q^2)^{-1}$ \cite{kroli00},
we show in figure \ref{fig1} the $C$ parameter versus $\sigma^2$ for various $P$ (note that $\beta$ changes along each curve,
because $\sigma^2$ varies) as calculated after the bound-state solutions of Eq.(\ref{psieq2bound}). As expected when $\sigma^2$ increases ($\sigma^2=0$
corresponds to the local case), the predicted fluctuation decreases.
Eq.(\ref{psieq2bound}) is also valid in the two-dimensional case for each transverse coordinate.

\begin{figure}
\includegraphics[width=0.50\textwidth]{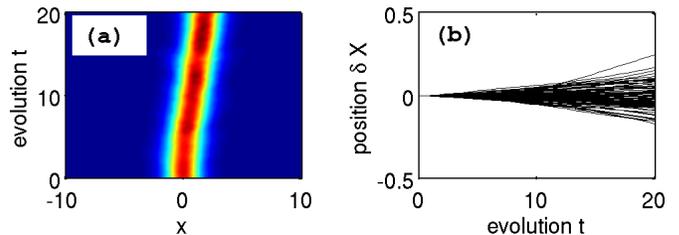}
\caption{(a) Typical dynamics of a solitary wave [Eq.(\ref{psieq2})] as obtained from
the numerical solutions of the stochastic equation (\ref{psieqpertu}) in the presence of 
randomness and exponential nonlocality ($P=6$, $\sigma^2=5$, $\eta_N=0.01$);
(b) Center of mass trajectories of the same bound-state ($P=6$, $\sigma^2=5$) for $100$
disorder realizations ($\eta_N=0.01$).
\label{fig2} }
\end{figure}

\begin{figure}
\includegraphics[width=0.50\textwidth]{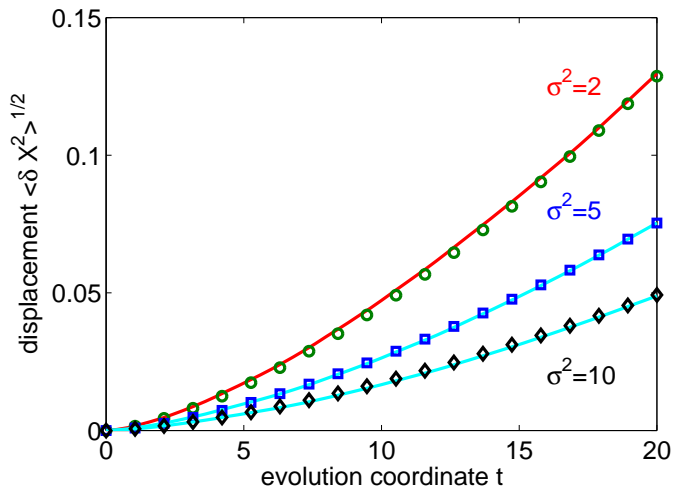}
\caption{(Color online) Comparison of the numerically (continuous lines, $P=6$, $\eta_N=0.01$) and theoretically 
(circles for $\sigma^2=2$, squares for $\sigma^2=5$ and diamonds for $\sigma^2=10$) calculated standard deviation of the
solitary wave position versus the evolution coordinate for three degrees of nonlocality and $100$ disorder realizations.}
\label{fig3} \end{figure}

To validate the previous analytical results, we resorted to the numerical integration of the 
stochastic partial differential equation resulting from a 1D exponential nonlocality; 
we adopted a pseudo-spectral stochastic Runge-Kutta method \cite{Werner97,Qiang00}.
Figure \ref{fig2}a shows a typical evolution starting from a bound state and displaying the random deviation of the SW. 
In figure \ref{fig2}b, we report various trajectories for a fixed SW power. Figure \ref{fig3} shows the 
calculated standard standard deviation for various degrees of nonlocality:
the analytical prediction after Eq.(\ref{ghnonlocal}) is indistinguishable from the numerical results.
%%% FIGURE  %%%%
%%% FIGURE  %%%%
%%% FIGURE %%%%

Before concluding, we consider the effect of non-paraxiality on 
the SW fluctuations. 
Ultra-thin nonlocal OSS were considered in \cite{ContiPRL04,Conti05PRL};
in this framework, non-paraxiality at the lowest order is described by the perturbation
$s=-\epsilon \psi_{xxxx}$,where $\epsilon$ is ratio between the
wavelength and the spatial beam waist (within some numerical constants, 
see \cite{Conti05PRL}).
Such a term is orthogonal to all the adjoint functions, 
with the exception of $\hat f_\theta$; this implies
that Eq.(\ref{ghnonlocal}) for the 
solitary-wave fluctuations still holds true in the ultra-focused regime, with
the addition of a linear increase of the soliton phase along
propagation: 
\begin{equation}
\delta\theta_{\text{non-paraxial}}=-t \frac{2\epsilon}{P'} \int u_\beta u_{xxxx} dx\text{,}
\end{equation}
corresponding to the perturbation to the nonlinear wave-vector
of the bound-state due to the non-paraxial term.

\noindent {\it Conclusions ---}
We have theoretically shown that nonlocality largely affects the dynamics of a solitary-wave in the presence of disorder;
this turns out into a random walk of the self-trapped beam position,
which is hampered by the filtering action of the nonlocal response, and ideally vanishes for a infinite degree of nonlocality. These results are expected to be specifically relevant for plasma-physics, Bose-Einstein condensates, thermal and thermo-diffusive media, liquid crystals and soft-colloidal matter, and suggest to employ highly nonlocal media for routing
information by solitons in order to moderate the effect of randomness. 

We acknowledge support from the CASPUR and CINECA
High Performance Computing initiatives.
The research leading to these results has received funding from the European
Research Council under the European Community's 
Seventh Framework Program (FP7/2007-2013)/ERC grant agreement n.201766.

%%%%%%%%%%%%

\end{document}